\documentclass{rspublic}
\usepackage{amsmath,paralist}
\usepackage{apacite}
\input{Qcircuit}

\DeclareMathOperator{\tr}{tr}

\DeclareMathOperator{\swt}{swt}
\DeclareMathOperator{\wt}{wt}

\newcommand{\C}{\mathbb{C}}
\newcommand{\F}{\mathbb{F}}
\newcommand{\nix}[1]{}
\newcommand{\sdual}{{\perp_s}}

\newcommand{\tdual}{{\perp_{t}}}

\newcommand{\ol}{\overline}
\newcommand{\floor}[1]{\lfloor #1\rfloor}

\newcommand{\scal}[2]{\langle #1\mid #2\rangle_s}

\renewcommand{\BOthers}{{\em et al}}
\renewcommand{\theBibCnt}{{\em\alph{BibCnt}\/}} %
\renewcommand{\BBAY}{{ }} %
\renewcommand{\BBAB}{\&}
\renewcommand{\BCBL}{{}}   
\renewcommand{\BCBT}{} 
\renewcommand{\bibliographytypesize}{\small} 

\begin{document}
\title[Subsystem Codes]{On Subsystem Codes Beating the Hamming or Singleton 
Bound}

\author[Klappenecker, Sarvepalli]{Andreas Klappenecker and Pradeep Kiran Sarvepalli}

\affiliation{Texas A\&M University, College Station, TX 77843, USA}

\label{firstpage}

\maketitle

\begin{abstract}{subsystem codes, operator codes, quantum Hamming bound, 
quantum Singleton bound} Subsystem codes are a generalization of
noiseless subsystems, decoherence free subspaces, and quantum
error-correcting codes. We prove a Singleton bound for $\F_q$-linear
subsystem codes.  It follows that no subsystem code over a prime field
can beat the Singleton bound.  On the other hand, we show the
remarkable fact that there exist impure subsystem codes beating the
Hamming bound.  A number of open problems concern the comparison in
performance of stabilizer and subsystem codes. One of the open
problems suggested by Poulin's work asks whether a subsystem code can
use fewer syndrome measurements than an optimal MDS stabilizer code
while encoding the same number of qudits and having the same
distance. We prove that linear subsystem codes cannot offer such an
improvement under complete decoding.
\end{abstract}

\section{Introduction}
Subsystem codes (sometimes also referred to as operator quantum
error-correcting codes) have emerged as an important new discovery in
the area of quantum error correcting codes, unifying the classes of
stabilizer codes, decoherence free subspaces and noiseless subsystems
\shortcite{bacon06a,knill06,kribs05,kribs06,kribs06b,poulin05}. 
From a practical perspective their importance lies in the fact that
they seem to offer better error recovery schemes than existing quantum
codes. 
Therefore, it is crucial to know under what circumstances these gains
can be attained and how to achieve them.

Recall that a quantum code $Q$ is a subspace in a finite dimensional
Hilbert space, $\mathcal{H}=\C^{q^n}$. A subsystem code is a quantum
code which can be further resolved into a tensor product {\em i.e.,}
$Q=A\otimes B$. Information is stored in system $A$, while system $B$,
referred to as the gauge subsystem, provides some additional
redundancy. By qudit we refer to a quantum bit with $q$ levels. We denote
the parameters of a subsystem code by $[[n,k,r,d]]_q$, indicating that
it is a $q$-ary code with length $n$, encodes $k$ qudits into the
subsystem $A$, and contains $r$ gauge qudits and has distance $d$.

Our goals in this paper are twofold. After reviewing the necessary
background on subsystem codes, we generalize the quantum Singleton
bound to $\F_{q}$-linear subsystem codes.  It follows that no
Clifford subsystem code over a prime field can beat the Singleton
bound. We use these results to show that if there exists an MDS
stabilizer code, then no linear subsystem code can outperform it in
the sense of requiring fewer syndrome measurements for error
correction.

\citeA{bacon06b} obtain a subsystem code
from two classical codes. We show that
this method is a special case of the Euclidean construction for subsystem
codes proposed in 
\shortciteA{aly06} and give a coding theoretic
analysis of these codes.

Since the early works on quantum error-correcting codes, it has been
suspected that impure codes should somehow perform better than the
pure codes. In particular, it was often conjectured that there might
exist impure quantum error-correcting codes beating the quantum
Hamming bound, but a proof remained elusive.  
\citeA{aly06} 
proved a Hamming bound for pure subsystem codes.
We show here that there exist impure subsystem
codes beating the Hamming bound.

\section{Background}\label{sec:background}
Let $\F_q$ be a finite field with $q$ elements and characteristic $p$. 
Let $C\subseteq \F_q^n$ be an $\F_q$-linear classical code denoted 
by $[n,k,d]_q$, where $k=\dim_{\F_q} C$ and $d$ is  the minimum distance of $C$.
We define $\wt(C)=\min \{\wt(c)\mid 0\neq c \in C \}=d$, where
$\wt(c)$ is the Hamming weight of $c$. Sometimes an alternative notation
$(n,K,d)_q$ is also used where $K=|C|$.
If $C$ is an $\F_p$-linear subspace over $\F_q$, then we say it is an
additive code.

If $x,y\in \F_q^n$, then their Euclidean inner product is defined as 
$x\cdot y= \sum_i x_iy_i$. The 
Euclidean dual of a code $C\subseteq \F_q^n$ is defined as 
$C^\perp = \{ y \in \F_q^n \mid x\cdot y=0 \mbox{ for all } x\in C\} $. 
We say that a code $C$ is self-orthogonal with respect to the Euclidean 
inner product if $C\subseteq C^\perp$.

We use the notation $(x|y)=(x_1,\ldots, x_n|y_1,\ldots,y_n)$ to 
denote concatenation of $x,y\in \F_q^n$.
Let $u=(a|b)$ and $v=(a'|b')$ be in $\F_q^{2n}$. 
We define the symplectic weight of $u$ as $\swt(u)=\{(a_i,b_i)\neq (0,0)\mid 1\leq i\leq n \}$
and the symplectic weight of a code $C\subseteq \F_q^{2n}$ as 
$\swt(C)=\min \{\swt(c)\mid 0\neq c \in C \}$.
For codes over 
$\F_q^{2n}$ another inner product plays a more important role in the context of 
quantum codes. 
 The  trace-symplectic product between $u,v$ is defined as
$\langle u|v\rangle_t = \langle(a|b)|(a'|b') \rangle_t =\tr_{q/p}(a'\cdot b-a\cdot b')$. 
The trace-symplectic dual of $C\subseteq \F_q^{2n}$ is defined as 
$C^\tdual = \{x\in \F_q^{2n} \mid \langle x|y\rangle_t=0, \mbox{ for all }y \in C\}$. 
If $C\subseteq C^\tdual$, we say that it is self-orthogonal with respect to the
trace-symplectic inner product.

\subsection{Subsystem codes from classical codes}
We now briefly review the background on subsystem codes. First we give
a group theoretic description and then give an alternate description
in terms of classical codes. Further details can be found in
\citeA{pre7,aly06}.

Let $q$ be the power of a prime $p$ and $\F_q$ a finite field with $q$
elements.  Let $B=\{ \ket{x}\mid x\in \F_q\}$ denote an orthonormal
basis for $\C^q$.  Let $X(a)$ and $Z(b)$ be unitary operators on
$\C^q$ whose action on any element $\ket{x}$ in $B$  is defined as
$$ X(a)\ket{x}=\ket{x+a} \mbox{ and }
Z(b)\ket{x}=\omega^{\tr_{q/p}(bx)}\ket{x},$$ where
$\omega=e^{j2\pi/p}$ is a primitive $p^{th}$ root of unity.  These
operators are a $q$-ary generalization of the well-known Pauli
matrices $X$ and $Z$.  Their action on an arbitrary element in $\C^q$
is obtained by invoking linearity.
Let $\mathcal{H}=\C^q\otimes \cdots \otimes
\C^q =\C^{q^n}$ and $\mathcal{E}$ be the error group on $\mathcal{H}$, defined
as the tensor product of $n$ such error operators {\em i.e.,}
$$ \mathcal{E}=\{ \omega^c E_1\otimes \cdots \otimes E_n\mid E_i = X(a_i)Z(b_i);
a_i,b_i\in \F_q; c\in \F_p\}.$$  The weight of an error $E = \omega^c E_1\otimes E_2\otimes \cdots \otimes E_n$ in $\mathcal{E}$
is defined as the number of $E_i$ which are not equal to identity and
it is denoted by $\wt(E)$. We can also associate to $E$ a vector
$\ol{E}=(a_1,\ldots,a_n|b_1,\ldots,b_n) \in \F_{q}^{2n}$. We define
the symplectic weight of $\ol{E}$ as
$$\swt(\ol{E})=|\{(a_i,b_i)\neq (0,0)\mid 1\leq i\leq n\}|=\wt(E).$$

Every nontrivial normal subgroup $N$ in $\mathcal{E}$ defines a subsystem code $Q$. 
Let $C_\mathcal{E}(N)$ be the centralizer of $N$ in $\mathcal{E}$ and $Z(N)$ the center of $N$. 
As a subspace the subsystem code $Q$ defined by $N$ is precisely
the same as the stabilizer code defined by $Z(N)$. 
By Theorem~4 in 
\citeA{pre7},
$Q$ can be decomposed as $A\otimes B$ where
$\dim B = |G : Z(N)|^{1/2}$ and
$$\dim A = |Z(\mathcal{E}) \cap G||\mathcal{E} : Z(\mathcal{E})|^{1/2}|N : Z(N)|^{1/2}/|N|.$$ 
Since information is stored only on subsystem $A$, we need only concern errors that affect $A$.
An error $E$ in $\mathcal{E}$ is detectable by subsystem $A$ if and only if $E$ is
contained in the set $\mathcal{E} - (NC_\mathcal{E}(N) - N)$. The distance of the code is
defined as
$$d=\min \{ \wt(E) \mid I \neq E\in NC_\mathcal{E}(N)-N\} =\wt(NC_\mathcal{E}(N)-N).$$ If
$NC_\mathcal{E}(N)=N$, then we define the distance of the code to be
$\wt(N)$. A distance $d$ subsystem code with $\dim A= K$, $\dim B=R$ is often denoted as
$((n,K,R,d))_q$ or $[[n,k,r,d]]_q$ if $K=q^k$ and $R=q^r$.
We say that $N$ is the \textsl{gauge group} of $Q$ and
$Z(N)$ its \textsl{stabilizer}.  The gauge group acts trivially on $A$.

In \citeA{pre7} 
we showed that
subsystem codes, much like the stabilizer codes, are related to the
classical codes over $\F_q^{2n}$ or $\F_{q^2}^n$, but with one
important difference. We no longer need the associated classical codes
to be self-orthogonal, thereby extending the class of quantum codes.
The gauge group $N$ can be mapped to a classical code $C$ over
$\F_{q}^{2n}$ and $C_\mathcal{E}(N)$ can be mapped to the trace-symplectic dual
of $C$. The following theorem \cite{pre7} 
shows how subsystem codes are related to classical codes .

\begin{theorem}\label{th:oqecfq}
Let $C$ be a classical additive subcode of\/ $\F_q^{2n}$ such that
$C\neq \{0\}$ and let $D$ denote its subcode $D=C\cap C^\tdual$. 
If $x=|C|$ and $y=|D|$, then
there exists an operator quantum error correcting code $C=
A\otimes B$ such that
\begin{compactenum}[i)]
\item $\dim A = q^n/(xy)^{1/2}$, 
\item $\dim B = (x/y)^{1/2}$.
\end{compactenum}
The minimum distance of subsystem $A$ is given by
\begin{compactenum}[(a)] 
\item $d=\swt((C+C^\tdual)-C)=\swt(D^\tdual-C)$ if\/ $D^\tdual \neq C$; 
\item $d=\swt(D^\tdual)$ if\/ $D^\tdual=C$. 
\end{compactenum}
Thus, the subsystem $A$ can
detect all errors in $\mathcal{E}$ of weight less than $d$, and can correct all
errors in $E$ of weight $\le \lfloor (d-1)/2\rfloor$.
\end{theorem}

We call codes constructed using theorem~\ref{th:oqecfq} as
\textsl{Clifford subsystem codes}.  Arguably, these codes cover the
most important subsystem codes, including the recently proposed
Bacon-Shor codes. In this paper, henceforth by a subsystem code
we will mean a Clifford subsystem code. 

A further simplification of the above construction is possible which takes
any pair of classical codes to give a subsystem code. We will just recall
the result here and study its application in the next section.
\begin{corollary}[Euclidean Construction]\label{th:cssoqec}
Let $X_i \subseteq \F_q^n$, be $[n,k_i]_q$ linear codes where $i\in \{1,2\}$. Then 
there exists an $[[n,k,r,d]]_q$   Clifford subsystem code with
\begin{compactitem}
\item $k=n-(k_1+k_2+k')/2$, 
\item $r=(k_1+k_2-k')/2$, and 
\item 
$d=\min \{ \wt((X_1^\perp\cap X_2)^\perp\setminus X_1), 
\wt((X_2^\perp\cap X_1)^\perp\setminus X_2) \}$,
\end{compactitem}
where $k'= \dim_{\F_q}(X_1\cap X_2^\perp)\times (X_1^\perp\cap X_2)$. 
\end{corollary} 

The result follows from Theorem~\ref{th:oqecfq} by defining
$C=X_1\times X_2$; it follows that $C^\sdual = X_2^\perp\times
X_1^\perp$ and $D=C\,\cap\,C^\sdual = (X_1\cap X_2^\perp)\times (X_2\cap
X_1^\perp)$, and the parameters are easily obtained from these definitions, 
see \citeA{aly06} for a detailed proof.

\subsection{Pure and impure subsystem codes}\label{ssec:impureCodes}
We can extend the notion of purity to subsystem codes also in a
straightforward manner. Let $N$ be the gauge group of a subsystem code
$Q$ with distance $d= \wt(C_\mathcal{E}(Z(N))-N)$. We say that $Q$ is
\textit{pure to $d'$} if there is no error of weight less than $d'$ in
$N$. The code is said to be \textit{exactly pure to $d'$} if $\wt(N)$
is $d'$ and it is said to pure if $d'\geq d$ .  The code is said to be
impure if it is exactly pure to $d'< d$.  This refinement to the
notion of purity was made in recognition of certain subtleties that
had to addressed when constructing other subsystem codes from existing
subsystem codes, see \citeA{aly06} 
for details.

In coding theoretic terms this can be translated as follows. 
Let $C$ be an additive subcode of $\F_{q}^{2n}$ and $D=C\cap
C^\tdual$.  By theorem~\ref{th:oqecfq}, we can obtain an
$((n,K,R,d))_q$ subsystem code $Q$ from $C$
that has minimum distance $d=\swt(D^\tdual - C)$.  
If $d'\leq \swt(C)$, then we say that
the associated operator quantum error correcting code is \textit{pure
to $d'$}. 

Extending these ideas of purity to subsystem codes is useful because it facilitates
the analysis of the parameters of the subsystem codes, as will become clear
when we derive bounds in the next section. If the codes are pure, then
it will be very easy to see that the subsystem code with the parameters $[[n,k,r,d]]_q$
satisfies $k+r\leq n-2d+2$. This is because then the subsystem code can also
be viewed as an $[[n,k+r,d]]_q$ stabilizer code, see theorem~11 in 
\citeA{aly06}
for  further details.

\section{Singleton upper bound for $\F_{q}$-linear subsystem codes}\label{sec:bounds}
\subsection{An upper bound for subsystem codes }\label{ssec:uBound}
We prove that the $\F_q$-linear subsystem codes with the parameters
$[[n,k,r,d]]_q$ satisfy a quantum Singleton like bound {\em viz.,} $k+r\leq n-2d+2$. It
will be seen that this reduces to the quantum Singleton bound if $r=0$. 
More interestingly, this reveals that there is a trade off in the size of 
subsystem $A$ and the gauge subsystem. One pays a price for the gains in error
recovery. The cost is the reduction in the information to be stored. 

Our proof for
this result is quite straightforward, though the intermediate details are a little
involved. First we show that a linear $[[n,k,r>0,d]]_q$ subsystem code that is 
exactly pure to 1 can be punctured to an $[[n-1,k,r-1,d]]_q$ code which retains the relationship
between $n,k,r,d$. If $d=2$ by repeated puncturing we either arrive at a pure code or a stabilizer code, both of which have upper bounds.
For $d>2$, two cases can arise, if the code is exactly pure to 1, we simply puncture it to get
a smaller code as in $d=2$ case. Otherwise, we puncture it
to get an $[[n-1,k,r+1,d-1]]_q$ code. By repeatedly shortening we either get a stabilizer code or a distance 2 code both of which have an upper bound. Keeping
track of the change in the parameters will give us an upper bound on the parameters of
the original code.

Let $w=(a_1,a_2,\ldots,a_n|b_1,b_2,\ldots,b_n)\in \F_q^{2n}$. We denote by 
$\rho(w) \in \F_q^{2n-2}$, the vector obtained by deleting
the first and the $n+1^{th}$ coordinates of $w$. Thus we have
$$\rho(w)=(a_2,\ldots,a_n|b_2,\ldots,b_n) \in \F_q^{2n-2}.$$
Similarly, given a classical code 
$C\subseteq \F_q^{2n}$ we denote  the puncturing of a codeword or code in the 
first and $n+1$ coordinates by $\rho(C)$.

For $\F_q$-linear codes instead of considering the trace symplectic inner product
we can consider the relatively simpler symplectic product. The symplectic product 
of $u=(a|b)$ and $v=(a'|b')$ in $\F_q^{2n}$ is defined as
$\langle u|v\rangle_s = \langle(a|b)|(a'|b') \rangle_s =a'\cdot b-a\cdot b'$. 
The symplectic dual of a code $C\subseteq \F_q^{2n}$ is defined as 
$C^\sdual = \{x\in \F_q^{2n} \mid \langle x|y\rangle_s=0, \mbox{ for all }y \in C\}$. 
It will be seen that $\langle u|v\rangle_t =\tr_{q/p}(\langle u|v \rangle_s)$. 

\begin{lemma}
Let $C \subseteq \F_{q}^{2n}$ be an $\F_q$-linear code with $(a|b) \in C$ and 
$(a'|b')\in C^\tdual$. Then $\langle (a|b)|(a'|b')\rangle_t=0$ if and only if 
$\langle (a|b)|(a'|b')\rangle_s=a\cdot b'-a'\cdot b=0$. 
It follows that $C^\tdual=C^\sdual$. 
\end{lemma}
\begin{proof}
If $\langle (a|b)|(a'|b')\rangle_s =0$, then 
$\tr_{q/p}(a'\cdot b-a\cdot b' )=0$. Since $C$ is linear $(\alpha a|\alpha b)$ is
also orthogonal to $(a'|b')$ for any $\alpha\in \F_q^\times$. Hence, 
$\tr_{q/p}(\alpha a'\cdot b-\alpha a\cdot b' )=0$. But $\tr$ is a nondegenerate
function. It follows that $a'\cdot b-a\cdot b'=0$. The converse is straightforward.
The equality of $C^\tdual=C^\sdual$ follows immediately from the first part
of the statement. 
\end{proof}

As we shall be concerned with $\F_q$-linear codes in this paper, we will 
focus only on the symplectic inner product in the rest of the paper.
\begin{lemma}\label{lm:xzBasis}
Let $C \subseteq \F_q^{2n}$ be an $\F_q$-linear code.
Then $C$ has an $\F_q$-linear basis of the form
$$B = \{ z_1,\ldots, z_k,z_{k+1},x_{k+1},z_{k+2},x_{k+2},\ldots, z_{k+r},x_{k+r} \}$$
where $\langle x_i |x_j \rangle_s  = 0 = \langle z_i |z_j \rangle_s$ and 
$\langle x_i| z_j\rangle_s = \delta_{i,j}$.
\end{lemma}
\begin{proof}
First we choose a basis $B=\{z_1,\ldots, z_k \}$ for a maximal isotropic subspace $C_0$ of $C$.
If $C_0\neq C$, then we can choose a codeword $x_1$ in $C$ that is orthogonal to all of the $z_k$ except one, say $z_1$ (renumbering if necessary).  We can scale $x_1$ by an element in $\F_q^\times$ so that 
$\langle z_1 |x_1\rangle_s=1$.  If $\langle C_0,x_1\rangle\neq C$, then we repeat the process  
until we have a basis of the desired form.
\end{proof}

\medskip
For the remainder of the section, we fix the following notation. 
By theorem~\ref{th:oqecfq}, we can associate with an 
$\F_q$-linear $[[n,k,r,d]]_q$ subsystem code two 
classical $\F_q$-linear codes
$C,D\subseteq \F_{q}^{2n}$ such that $D=C\cap C^\sdual$,
$|C|=q^{n-k+r}$, $|D|=q^{n-k-r}$ and $\swt(D^\sdual\setminus C)=d$.  
By lemma~\ref{lm:xzBasis}, we can also
assume that $C$ is generated by
$$C =\langle z_1,\ldots, z_s,z_{s+1},x_{s+1},\ldots, z_{s+r},x_{s+r}\rangle,$$ 
where $s=n-k-r$
and the vectors $x_i$, $z_i$ in $\F_q^{2n}$ satisfy the relations 
$\langle x_i|x_j \rangle_s =0 = \langle z_i|z_j\rangle_s$ and
$\langle x_i|z_j \rangle_s=\delta_{i,j}$. These relations on $x_i,z_i$  imply that 
\begin{eqnarray*}
C^\sdual &=&\langle z_1,\ldots, z_s,z_{s+r+1},x_{s+r+1},\ldots, z_{s+r+k},x_{s+r+k}\rangle,\\
D=C\cap C^\sdual &=&\langle z_1,\ldots, z_s \rangle,\\ 
D^\sdual &=& \langle z_1,\ldots, z_s,z_{s+1},x_{s+1},\ldots, z_{n},x_{n}\rangle.\end{eqnarray*}

\begin{lemma}\label{lm:punc1lin}
An $\F_{q}$-linear $[[n,k,r>0,d\geq 2]]_q$ Clifford subsystem code exactly pure to $1$ can be punctured to 
an $\F_{q}$-linear $[[n-1,k,r-1, \geq d]]_q$ code. 
\end{lemma}
\begin{proof}
As mentioned above, we can associate to the subsystem code two classical codes 
$C,D\subseteq \F_q^{2n}$. 
Two cases arise depending on $\swt(D)$. 
\begin{compactenum}
\item[a)] If $\swt(D)=1$, then without loss of generality we can assume that $\swt(z_1)=1$.
Further, $z_1$ can be taken to be of the form $(1,0,\ldots,0|a,0,\ldots,0)$. 
and  for $i\neq 1$, because of $\F_q$-linearity of the codes we can choose every $x_i,z_i$ to be of the form 
$(0,a_2,\ldots,a_n|b_1,b_2,\ldots,b_n)$. 
Further, as $x_i,z_i$ must satisfy the orthogonality relations 
with $z_1$ {\em viz.,} $\langle z_1|z_i \rangle_s=0= \langle z_1|x_i \rangle_s$, for 
$i>1$ we can choose $x_i,z_i$ to be of the form $(0,a_2,\ldots,a_n|0,b_2,\ldots,b_n)$.
It follows that because of the form of $x_i$ and $z_i$ puncturing the first and $n+1^{th}$ coordinate
will not alter these orthogonality relations, in particular $\langle \rho(x_i)|\rho(z_i)\rangle_s\neq 0$
for ${s+1}\leq i\leq n$. 

Letting $\rho(x_i)=x_i'$, $\rho(z_i)=z_i'$ and observing that
$\rho(z_1)=(0,\ldots,0|0,\ldots,0)$, we see that the code 
$\rho(C) = \langle z_2',\ldots, z_s', z_{s+1}',x_{s+1}',\ldots,z_{s+r}',x_{s+r}'\rangle$.
Denoting by $D_p=\rho(C)\cap \rho(C)^\sdual$ it is immediate that $D_p$ is generated by
$\{ z_2',\ldots, z_s'\}$ while 
$D_p^\sdual= \langle z_2',\ldots, z_s', z_{s+1}',x_{s+1}',\ldots,z_{n}',x_{n}'\rangle$.
Hence $\rho(C)$ defines an $[[n-1,k,r,\swt(D_p^\sdual\setminus \rho(C))]]_q$ code. 

Next we show that $\swt(D_p^\sdual\setminus \rho(C))\geq d$. 
Let $u=(a_2,\ldots,a_n|b_2,\ldots, b_n)$ be in
$D_p^\sdual \setminus \rho(C)$, then we can easily verify that 
$(0,a_2,\ldots,a_n|0,b_2,\ldots, b_n)$ is orthogonal to all
$z_i$, $1\leq i\leq s$ and hence it is in $D^\sdual$. It cannot be in $C$ as that would imply
that $u$ is in $\rho(C)$. But $\swt(D^\sdual \setminus C)\geq d$. Therefore $\swt(u)\geq d$.
and $\rho(C)$ defines an $[[n-1,k,r,\geq d]]_q$ code.
By choosing $C'= \langle z_2',\ldots, z_s',z_{s+1}',z_{s+2}',x_{s+2}',\ldots, z_{s+r}',x_{s+r}'\rangle$
we can conclude that there exists an $[[n-,k,r-1,d]]_q$ code. Alternatively, apply 
theorem~16 in 
\citeA{aly06}.
\item[b)] If $\swt(D) >1$, then we can assume that $\swt(z_{s+1})=1$ and form the code 
$C'= \langle z_1,\ldots, z_s,z_{s+1},z_{s+2},x_{s+2},\ldots, z_{s+r},x_{s+r}\rangle$. It is clear that 
$C'$ defines an $[[n,k,r-1,d]]_q$ code that is pure to $1$ with $\swt(C'\cap C'^\sdual)=1$. But this is just
the previous case, from which we can conclude that there exists an $[[n-1,k,r-1,\geq d]]_q$ code.
\end{compactenum}
\end{proof}
Lemma~\ref{lm:punc1lin} allows us to establish a bound for distance 2 codes which can then be
used to prove the bound for arbitrary distances. 
\begin{lemma}\label{lm:boundlind2}
An impure $\F_{q}$-linear $[[n,k,r,d=2]]_q$ Clifford subsystem code satisfies
$$ 
k+r\leq n-2d+2.
$$
\end{lemma}
\begin{proof}
Suppose that there exists an $\F_{q}$-linear $[[n,k,r,d=2]]_q$ impure
subsystem code such that $k+r> n-2d+2$; in particular, this code must
be pure to~$1$. By lemma~\ref{lm:punc1lin} it can be punctured to give
an $[[n-1,k,r-1,\geq d]]_2$ subsystem code. If this code is pure, then
$k+r-1 \leq n-1-2d+2$ holds, contradicting our assumption $k+r>
n-2d+2$; hence, the resulting code is once again impure and pure to 1.

Now we repeatedly apply lemma~\ref{lm:punc1lin} to puncture the
shortened codes until we get an $[[n-r,k,0,\geq d]]_q$ subsystem
code. But this is a stabilizer code which must obey the Singleton bound 
$k\leq n-r-2d+2$, contradicting our initial assumption 
$k+r>n-2d+2$. Therefore, we can conclude that $k+r\leq n-2d+2$.
\end{proof}

If the codes are of distance greater than 2, then we puncture the code
until it either has distance 2 or it is a pure code. The following
result tells us how the parameters of the subsystem codes vary on
puncturing.
\begin{lemma}\label{lm:punc1linp2}
An impure $\F_{q}$-linear $[[n,k,r,d\geq 3]]_q$  Clifford subsystem code exactly
pure to $d'\geq 2$ implies the existence of an $\F_{q}$-linear
$[[n-1,k,r+1,\geq d-1]]_q$ subsystem code.
\end{lemma}
\begin{proof}
Recall that the existence of an $[[n,k,r,d\ge 3]]_q$ subsystem code
implies the existence of $\F_q$-linear codes $C$ and $D$ such that 
$$C=\langle z_1,\ldots, z_s,z_{s+1},x_{s+1},\ldots,
z_{s+r},x_{s+r}\rangle,$$ with $s=n-k-r$, and $D=C\cap C^\sdual$, see above. 

The stabilizer code defined by $D$ satisfies $k+r=n-s\leq n-2d+2$, or
equivalently $s\geq 2d-2$; it follows that $s\ge 2$, since $d\ge d'\ge
2$. Without loss of generality, we can take $z_1$ to be of the form
$(1,a_2,\ldots,a_n|b_1,b_2\ldots,b_n)$ for if no such codeword exists
in $D$, then $(0,0,\ldots,0|1,0,\ldots,0)$ is contained in $D^\sdual$,
contradicting the fact that $\swt(D^\sdual)\geq 2$.  Consequently, we
can choose $z_2$ in $D$ to be of the form
$(0,c_2,\ldots,c_n|1,d_2,\ldots,d_n)$, and we may further assume that
$b_1=0$ in $z_1$. The form of $z_1$ and $z_2$ allows us to assume
that any remaining generator of $C$ is of the form 
$(0,u_2,\ldots,u_n|0,v_2,\ldots,v_n)$.  

Let $\rho$ be the map defined by puncturing the first and $(n+1)^{th}$
coordinate of a vector in $C$. Define for all $i$ the punctured
vectors $x_i'=\rho(x_i)$ and $z_i'=\rho(z_i)$. Then one easily checks
that $\scal{\rho(x_i)}{\rho(x_j)}=0=\scal{\rho(z_i)}{\rho(z_j)}$ for
all indices $i$ and $j$, and
$\scal{\rho(x_i)}{\rho(z_j)}=\delta_{i,j}$ if $i\ge s+1$ or $j\ge 3$,
and that $\scal{\rho(z_1)}{\rho(z_2)}=-1$.

Let us look at the punctured code $\rho(C)$,
$$\rho(C)=\langle z_3',\ldots,z_s',z_{s+1}',x_{s+1}',\ldots,
z_{s+r}',x_{s+r}',z_{1}',z_{2}'\rangle.$$ Since
$\scal{\rho(z_1)}{\rho(z_2)}=-1$ we have $D_p=\rho(C)\cap
\rho(C)^\sdual =\langle z_3',\ldots,z_s'\rangle $, whence
$|D_p|=|D|/q^2$.  As $\swt(C)\geq 2$, it follows that $|\rho(C)|=|C|$.
Thus $\rho(C)$ defines an $[[n-1,k,r+1,\swt(D_p^\sdual\setminus
\rho(C))]]_q$ subsystem code.  

Recall that the code $D$ is generated by $s\ge 2$ vectors; we will
show next that our assumptions actually force $s\ge 3$.  Indeed, if
$s=2$, then $|D|=q^2$ and $|D^\sdual|=q^{2n-2}$. Under the assumption
$\swt(D^\sdual)\geq 2$, it follows that $|\rho(D^\sdual)|
=|D^\sdual|=q^{2n-2}$.  But as $\rho(D^\sdual) \subseteq \F_q^{2n-2}$
this implies that $\rho(D^\sdual)=\F_q^{2n-2}$.  Since $\F_{q}^{2n-2}$
has $2n-2$ independent codewords of symplectic weight one, $D^\sdual$
must have $2n-2$ independent codewords of symplectic weight
two. However, this contradicts our assumptions on the minimum distance
of the subsystem code: 
\begin{compactenum}[(a)]
\item If $C$ is a proper subspace of $D^\sdual$, then the minimum
distance $d$ is given by $d=\swt(D^\sdual\setminus C)\ge 3$; thus, the
weight 2 vectors must all be contained in $C$, which shows that
$|C|=q^{2n-2}=|D|$, contradicting $|C|<|D^\sdual|$.
\item If $C=D^\sdual$, then the minimum distance is given by
$d=\swt(D^\sdual)=2$, contradicting our assumption that $d\ge 3$. 
\end{compactenum}
Thus, from now on, we can assume that $s\geq 3$.

Before bounding the minimum distance of the punctured subsystem code,
we are going to show that $D_p^\sdual=\rho(D^\sdual)$.  Let
$w=(u_1,u_2,\ldots,u_n|v_1,v_2,\ldots,v_n)$ be a vector in
$D^\sdual$. For $3\leq i\leq s$, the vectors $z_i$ are of the
form $(0,a_2,\ldots,a_n|0,b_2,\ldots,b_n)$;
thus, it follows from $\langle w|z_i\rangle_s=0$ that $\langle
\rho(w)|z_i'\rangle_s=0$.  Hence $\rho(w)$ is in $D_p^\sdual$, which
implies $\rho(D^\sdual) \subseteq D_p^\sdual$. We have $|D_p^\sdual|
=q^{2n-2}/|D_p|=q^{2n}/|D|=|D^\sdual|$, and we note that
$|D^\sdual|=|\rho(D^\sdual)|$, because $\swt(D^\sdual)\geq 2$; hence,
$D_p^\sdual=\rho(D^\sdual)$.

Let $w'=(u_2,\ldots,u_n|v_2,\ldots,v_n)$ be an arbitrary vector in
$\rho(D^\sdual) \setminus \rho(C)$.  It follows that there exist some
$\alpha, \beta$ in $\F_q$ such that
$w=(\alpha,u_2,\ldots,u_n|\beta,v_2,\ldots,v_n)$ is in $D^\sdual;$ it is
clear that $w$ cannot be in $C$, since then $\rho(w)=w'$ would be in
$\rho(C)$; hence, $\swt(w)\geq d$. It immediately follows that 
$\swt(D_p^\sdual\setminus \rho(C))\geq d-1$.  Hence $\rho(C)$ defines
an $[[n-1,k,r+1,\geq d-1]]_q$ subsystem code.
\end{proof}

Now we are ready the prove the upper bound for an arbitrary subsystem code.
Essentially we reduce it to a pure code or distance two code by repeated puncturing
and bound the parameters by carefully tracing the changes.
\begin{theorem}\label{th:boundlin}
An $\F_{q}$-linear $[[n,k,r,d\geq 2]]_q$ Clifford subsystem code satisfies 
\begin{eqnarray}k+r\leq n-2d+2.\label{eq:singBound}\end{eqnarray}
\end{theorem}
\begin{proof}
The bound holds for all pure codes, see 
\citeA{aly06}. So
assume that the code is impure.  If $d=2$, then the relation holds by
lemma~\ref{lm:boundlind2}; so let $d\geq 3$. If the code is exactly pure to $1$,
then it can be punctured using lemma~\ref{lm:punc1lin} to give an
$[[n-1,k,r-1,d'=d]]_q$ code, otherwise it can be punctured using
lemma~\ref{lm:punc1linp2} to obtain an $[[n-1,k,r+1,d'\geq d-1]]_q$
code. If the punctured code is pure, then it follows that either 
$k+r-1\leq n-1-2d+2$ or $k+r+1\leq n-1-2d'+2\leq n-1-2(d-1)+2$ holds; 
in both cases, these inequalities imply that $k+r\leq n-2d+2$.

If the resulting code is impure, then if it is exactly pure 
to $1$ we puncture the code again using lemma~\ref{lm:punc1lin}, 
if not we puncture using lemma~\ref{lm:punc1linp2}, until we get a
pure code or a code with distance two. Assume that we punctured $i$ times
using lemma~\ref{lm:punc1lin} and $j$ times using
lemma~\ref{lm:punc1linp2}, then the resulting code is an
$[[n-i-j,k,r+j-i,d'\geq d-j]]_q$ subsystem code. Since pure subsystem
codes and distance 2 subsystem codes satisfy $$k+r+j-i\leq
n-i-j-2d'+2\leq n-i-j-2(d-j)+2,$$ it follows that $k+r\leq n-2d+2$ holds. 
\end{proof}

\medskip
When the subsystem codes are over a prime alphabet, this bound holds
for all codes over that alphabet.  In the more general case where the
code is not linear, numerical evidence indicates that it is unlikely
that the additive subsystem codes have a different bound. We have
shown that a large class of impure codes already satisfy this bound.
We conjecture that all subsystem codes satisfy $k+r\leq n-2d+2$. Next,
we give an application of this upper bound.

\subsection{Can subsystem codes improve upon MDS stabilizer codes? }
\label{ssec:mds}

In this subsection, we compare stabilizer codes with subsystem codes.
We first need to establish the criteria for the comparison, since
subsystem codes cannot be universally better than stabilizer codes.
For example, it is known that an $[[n,k,r,d]]_q$ subsystem code can be
converted to an $[[n,k,d]]_q$ stabilizer code (see \citeA{aly06},
lemma~10 for a proof of this claim); this implies that no
$[[n,k,r,d]]_q$ subsystem code can beat an optimal $[[n,k,d']]_q$
stabilizer code in terms of minimum distance, as $d'\geq d$.  One of
the attractive features of subsystem codes is a potential reduction of
the number of syndrome measurements, and we use this criterion as the
basis for our comparison.

First, we must highlight a subtle point on the required number of
syndrome bits for an $\F_q$-linear $[n,k,d]_q$ code. A complete decoder,
will require $n-k$ syndrome bits. Complete decoders are also optimal
decoders. A bounded
distance decoder on the other hand can potentially decode with 
fewer syndrome bits. Bounded distance decoders typically 
decode  up to $\floor{(d-1)/2}$.
However, to the best of our knowledge, except for
the lookup table decoding method, all bounded distance decoders
also require $n-k$ syndrome bits.
As the complexity of decoding using a  lookup table increases exponentially 
in $n-k$ it is  highly impractical for long lengths. We therefore assume
that for practical purposes, that we need  $n-k$ syndrome bits.

Similarly, for an $\F_q$-linear $[[n,k,r,d]]_q$ subsystem code, a complete 
decoder will require $n-k-r$ syndrome measurements, as is shown 
in~\ref{sec:appendix}. We are not aware of any quantum code,
stabilizer or subsystem, for which there exists a bounded distance
decoder that uses less than $n-k-r$ syndrome measurements to
perform bounded distance decoding.
The work by \citeA{poulin05} prompts the following question: Given an
optimal $[[k+2d-2,k,d]]_q$ MDS stabilizer code, is it possible to find
an $[[n,k,r,d]]_q$ subsystem code that uses fewer syndrome
measurements?

There exist numerous known examples of subsystem codes that improve
upon nonoptimal stabilizer codes. The fact that the stabilizer code is
assumed to be optimal makes this question interesting.  The Singleton
bound $k+r\le n-2d+2$ of an $\F_q$-linear $[[n,k,r,d]]_q$ subsystem
code implies that the number $n-k-r$ of syndrome measurements is
bounded by $n-k-r\geq 2d-2$; thus, for fixed minimum distance $d$,
there exists a trade off between the dimension $k$ and the difference
$n-r$ between length and number of gauge qudits.

\begin{corollary}\label{th:betterQMDS} Under complete decoding
an $\F_q$-linear $[[n,k,r,d\geq 2]]_q$ Clifford subsystem code cannot use
fewer syndrome measurements than an $\F_q$-linear $[[k+2d-2,k,d]]_q$
stabilizer code.
\end{corollary}
\begin{proof}
Seeking a contradiction, we assume that there exists an
$[[n,k,r,d]]_q$ subsystem code that requires fewer syndrome
measurements that the optimal $[[k+2d-2,k,d]]_q$ MDS stabilizer code.
In other words, the number of syndrome measurement yield the
inequality $k+2d-2-k> n-k-r$, which is equivalent to $k+r > n-2d+2$,
but this contradicts the Singleton bound.
\end{proof}

\citeA{poulin05} showed by exhaustive computer search that there does
not exist an $[[5,1,r>0,3]]_2$ subsystem code. The above result
confirms his computer search and shows further that not even allowing
longer lengths and more gauge qudits can help in reducing the number
of syndrome measurements. In fact,
we conjecture that corollary~\ref{th:betterQMDS} holds for bounded distance
decoders also.

We wish to caution the reader that gains in error recovery cannot be
quantified purely by the number of syndrome measurements.  In practice,
more complex measures such as the simplicity of the decoding algorithm
or the resulting threshold in fault-tolerant quantum computing are
more relevant. The drawback is that the comparison of large classes of
codes becomes unwieldy when such complex criteria are used. 

\section{Subsystem codes on a lattice}
Bacon gave the first family of subsystem codes generalizing the ideas
of Shor's $[[9,1,3]]_2$ code \cite{bacon06a}.  Recently, he and
Casaccino gave another construction which generalizes this further by
considering a pair of classical codes \cite{bacon06b}.  We show that
this method is a special case of theorem~\ref{th:oqecfq}. Since this
construction is not limited to binary codes and our proofs remain
essentially the same, we will immediately discuss a generalization to
nonbinary alphabets.

\begin{theorem}\label{th:latticeCodes}
For $i\in \{1,2\}$, let $C_i \subseteq \F_q^{n_i}$ be $\F_q$-linear
codes with the parameters $[n_i,k_i,d_i]_q$. Then there exists a
Clifford subsystem code with the parameters
$$[[n_1n_2, k_1k_2, (n_1-k_1)(n_2-k_2), \min\{d_1,d_2 \}]]_q$$
that is pure to $d_p=\min\{d_1^\perp,d_2^\perp\}$, where $d_i^\perp$ denotes the minimum distance of $C_i^\perp$. 
\end{theorem}

\begin{proof}
Let $C$ be the classical linear code given by $C=(\F_q^{n_1}\otimes
C_2^\perp) \times (C_1^\perp \otimes \F_q^{n_2})$.  Then $\dim C =
n_1(n_2-k_2)+n_2(n_1-k_1)$ and $\swt(C\setminus \{0\})\ge
\min\{d_1^\perp,d_2^\perp\}$.  The symplectic dual of $C$ is given by
\begin{eqnarray*}
C^\sdual &=&  (C_1^\perp\otimes \F_q^{n_2})^\perp \times (\F_q^{n_1}\otimes C_2^\perp)^\perp\\
&=& (C_1\otimes \F_q^{n_2}) \times (\F_q^{n_1}\otimes C_2).
\end{eqnarray*}
We have $\dim C^\sdual= k_1n_2+n_1k_2$. 
The code $D=C\cap C^\sdual$ is given by  
\begin{eqnarray*}
\begin{split}
D &=\left((\F_q^{n_1}\otimes C_2^\perp) \times (C_1^\perp \otimes \F_q^{n_2}) \right)\cap
\left((C_1\otimes \F_q^{n_2}) \times (\F_q^{n_1}\otimes C_2) \right)\\
&=\left((\F_q^{n_1}\otimes C_2^\perp) \cap (C_1\otimes \F_q^{n_2}) \right)\times
\left(  (C_1^\perp \otimes \F_q^{n_2})\cap (\F_q^{n_1}\otimes C_2) \right)\\
&= (C_1\otimes C_2^\perp) \times (C_1^\perp \otimes C_2), 
\end{split}
\end{eqnarray*}
and $\dim D = k_1(n_2-k_2)+k_2(n_1-k_1)$.  It follows that $\dim C -
\dim D = 2(n_1-k_1)(n_2-k_2)$ and $\dim C^\sdual -\dim D =
2k_1k_2$. Using corollary~\ref{th:cssoqec}, we can get a subsystem
code with the parameters $$[[n_1n_2, k_1k_2, (n_1-k_1)(n_2-k_2),
d=\swt(D^\sdual\setminus C)]]_q $$ that is pure to $d_p=\min\{d_1^\perp,d_2^\perp\}$. It remains to show
that $d=\min\{d_1,d_2\}$. 

Since $D= (C_1\otimes
C_2^\perp) \times (C_1^\perp \otimes C_2)$, we have
\begin{eqnarray*}
D^\sdual &=&  (C_1^\perp \otimes C_2)^\perp \times (C_1\otimes C_2^\perp)^\perp\\
&=&  
\left((C_1 \otimes \F_q^{n_2})  + (\F_q^{n_1}\otimes C_2^\perp)\right)
\times 
\left((\F_q^{n_1}\otimes C_2) + 
(C_1^\perp\otimes \F_q^{n_2})\right).
\end{eqnarray*}
In the last equality, we used the fact that vectors $u_1\otimes
u_2$ and $v_1\otimes v_2$ are orthogonal if and only if $u_1\perp v_1$
or $u_2\perp v_2$.

For $i\in \{1,2\}$, let $G_i$ and $H_i$ respectively denote the
generator and parity check matrix of the code $C_i$. Without loss of
generality, we may assume that these matrices are in standard form
$$
H_i=\left[\begin{array}{cc}I_{n_i-k_i} & P_i\end{array}\right] \mbox{ and }  
G_i=\left[\begin{array}{cc}-P_i^t &I_{k_i} \end{array} \right],
$$
where $P_i^t$ is the transpose of $P_i$. Let $H_i^c=\left[\begin{array}{cc} 0& I_{k_i} \end{array}\right]$. Using these notations, the generator matrices of $C$ and $D^\sdual$ can be written as 
$$
G_C=\left[\begin{array}{cc}I_{n_1}\otimes H_2 &0\\ 0&H_1\otimes I_{n_2} \end{array}\right]\quad \text{and} \quad 
G_{D^\sdual} =  
\left[ \begin{array}{cc}
G_1 \otimes H_{2}^c &0 \\I_{n_1}\otimes H_2&0\\
0&H_{1}^c\otimes G_2\\0&H_1\otimes I_{n_2}
\end{array}\right].
$$ 
It follows that the minimum distance $d$ is given by 
\begin{eqnarray*}
\begin{aligned}
\swt(D^\sdual\setminus C)=\min&\left\{\wt\left(\left\langle  \begin{array}{c}
G_1 \otimes H_{2}^c \\I_{n_1}\otimes H_2
\end{array}\right\rangle \setminus \left\langle \begin{array}{c}
I_{n_1}\otimes H_2
\end{array} \right\rangle \right)\right.,\\
&\;\;\left. \wt\left(\left\langle  \begin{array}{c}
H_{1}^c\otimes G_2\\H_1\otimes I_{n_2}
\end{array} \right\rangle\setminus  \left\langle\begin{array}{c}
H_1\otimes I_{n_2}
\end{array} \right\rangle\right) \right\}.
\end{aligned}
\end{eqnarray*}
Let us compute 
$$\wt\left(\left\langle  \begin{array}{c}
H_{1}^c\otimes G_2\\H_1\otimes I_{n_2}
\end{array} \right\rangle\setminus  \left\langle\begin{array}{c}
H_1\otimes I_{n_2}
\end{array} \right\rangle\right).$$
If minimum weight codeword is present in $D^\sdual\setminus C$, it must be expressed as linear 
combination of at least one row from $\left[H_1^c\otimes G_2 \right]$ otherwise the 
codeword is entirely in $C$. Recall that $ H_1=[\begin{array}{cc}I_{n_1-k_1} & P_1\end{array}]$ and 
$H_1^c=[ \begin{array}{cc}0 &I_{k_1}\end{array}]$. Letting $P_1=(p_{ij})$,
we can write  
\begin{eqnarray*}
\left[  \begin{array}{c}
H_{1}^c\otimes G_2\\H_1\otimes I_{n_2}
\end{array} \right] = \left[\begin{array}{cccccccc}
0&0& \dots&0&G_2&0\\
0&0& \dots&0&0&G_2&0\\
\dots & \dots&\dots&\dots &\dots&\dots&\dots&\dots\\
0&0&\dots& 0&0&\dots&\dots&G_2\\\hline
I_{n_2}&0&\dots&0&p_{11}I_{n_2}&\dots&\dots&p_{1k_1}I_{n_2}\\
0&I_{n_2}&\dots&\dots&p_{21}I_{n_2}&\dots&\dots&p_{2k_1}I_{n_2}\\
\dots & \dots&\dots&\dots &\dots&\dots&\dots&\dots\\
0&0&\dots&I_{n_2}&p_{(n_1-k_1)1}I_{n_2}&\dots&\dots&p_{(n_1-k_1)k_1}I_{n_2}
\end{array} \right].
\end{eqnarray*}
Now observe that any row below the line in the above matrix can has a weight of only one in each of the last
$k_1$ blocks of size $n_2$. And any linear combination of them involving less than $d_2$ 
and at least one generator from the rows above must have a weight $\geq d_2$. If on the other hand there are more than
$d_2$ rows involved, then the first $n_2(n_1-k_1)$ columns will have a weight $\geq d_2$. Thus in either
case the weight of an element that involves a generator from $\left[ H_1^c\otimes G_2\right]$ must have a weight $ \geq d_2$. On the other hand, the minimum weight of the span of $\left[H_1^c\otimes G_2 \right]$ is 
$\wt(C_2)=d_2$, from which we can conclude that
$$\wt\left(\left\langle  \begin{array}{c}
H_{1}^c\otimes G_2\\H_1\otimes I_{n_2}
\end{array} \right\rangle\setminus  \left\langle\begin{array}{c}
H_1\otimes I_{n_2}
\end{array} \right\rangle\right)=d_2.$$
Because of the symmetry in the code we can argue that 
$$\wt\left(\left\langle  \begin{array}{c}
G_1 \otimes H_{2}^c \\I_{n_1}\otimes H_2
\end{array}\right\rangle \setminus \left\langle \begin{array}{c}
I_{n_1}\otimes H_2
\end{array} \right\rangle \right)=d_1$$
and consequently $d=\min\{ d_1,d_2\}$, which proves the theorem.
\end{proof}

\subsection{Bacon-Shor codes}
\citeA{bacon06a} proposed one of the first families of subsystem codes based on square
lattices.  
A trivial modification using rectangular lattices instead of square
ones gives the following codes, see also \citeA{bacon06b}. The relevance of these codes will be
seen later in \S\ref{sec:packing}. Using the same notation as in 
theorem~\ref{th:latticeCodes}, let $G_i=[1,\ldots,1]_{1\times i}$
and $H_i$ be the matrix defined as
$$
H_i=\left[ \begin{array}{ccccccc}
1&1&&&&&\\&1&1&&&&\\
&& &\ddots &&&\\
&&&&1&1&\\&&&&&1&1
 \end{array}\right]_{i-1\times i}
$$
and $C$, the additive code generated by the following matrix.
$$
G= \left[\begin{array}{cc}
I_{n_1}\otimes H_{n_2}&0\\
0&H_{n_1}\otimes I_{n_2}
 \end{array}\right].
$$
Observe that $G_i$ generates an $[i,1,i]_q$ code with distance $i$. 
By theorem~\ref{th:latticeCodes}, $G_{n_1}$ and $G_{n_2}$ will give us the 
following family of codes 
\begin{corollary}\label{co:rectLattice}
There exist $[[n_1n_2,1,(n_1-1)(n_2-1),\min\{n_1,n_2\}]]_q$ Clifford subsystem codes.
\end{corollary}

\section{Subsystem codes and packing}\label{sec:packing}
We investigate whether subsystem codes lead to better codes because of the 
decomposition of the code space. Since the early days of quantum codes,
it has recognized that the degeneracy of quantum codes could lead to 
a more efficient quantum code and allow for a much more compact 
packing of the subspaces in the Hilbert space. But so far it has not
been shown for stabilizer codes. We can derive  similar bound for 
subsystem codes. 
\citeA{aly06} 
showed the following theorem for pure subsystem codes.
\begin{theorem}
A pure $((n,K,R,d))_q$ Clifford subsystem code satisfies 
\begin{eqnarray} 
\sum_{j=0}^{\lfloor (d-1)/2\rfloor}\binom{n}{j} (q^2-1)^j \leq q^n/KR.
\end{eqnarray}
\end{theorem}
It is natural to ask if impure subsystem codes also satisfy this bound. We show 
that they do not by giving an explicit counterexample. This counter example comes 
from the codes proposed by 
\citeA{bacon06a}. 
Recall the Bacon-Shor codes are $[[n^2,1,(n-1)^2,n]]_2$ subsystem codes. The $[[9,1,4,3]]_2$ 
is an interesting code. 
We can check that it satisfies the Singleton bound for subsystem codes as
$$ k+r=1+4 = n-2d+2=9-6+2.$$ So it is an optimal code. 
More interestingly, substituting the 
parameters of the $[[9,1,4,3]]_2$ Bacon-Shor code in the above inequality we get
$$
\sum_{j=0}^{1}\binom{9}{j}3^j = 28 >  2^{9-5}=16.
$$
Therefore the $[[9,1,4,3]]_2$ Bacon-Shor code beats the quantum Hamming bound
for the pure subsystem codes proving the following result.

\begin{theorem}
There exist impure $((n,K,R,d))_q$ Clifford subsystem codes that do not satisfy 
$$ 
\sum_{j=0}^{\lfloor (d-1)/2\rfloor}\binom{n}{j} (q^2-1)^j \leq q^n/KR.
$$
\end{theorem}

An obvious question is why impure codes can potentially pack more efficiently
than the pure codes. Let us understand this by looking at the $[[9,1,4,3]]_2$
code a little more closely. This code encodes information into a subspace, $Q$
where  $\dim Q= 2^{k+r}=2^5$. As it is a subsystem code $Q$ can be decomposed as $Q=A\otimes B$,
with $\dim A =2^k=2$ and $\dim B= 2^r=2^4$. In a pure single error 
correcting code all single errors must take the code space into orthogonal 
subspaces. In an impure code this is not required two or more distinct errors
can take the code space to the same orthogonal space. In the Bacon-Shor code
a phase flip error on any of the first three qubits will take the 
code space to same orthogonal subspace and because of this we cannot distinguish
between these errors. However, it is not a problem because 
we can restore the code space with respect to $A$
even though we cannot restore $B$. Thus instead of requiring $9$ orthogonal subspaces
as in a pure code, we only require 3 orthogonal subspaces to correct for 
any single phase flip error. Considering the bit flip errors and the combinations we need only
$9$ orthogonal subspaces. Thus with the original code space this means we need to pack
ten $2^5$-dimensional subspaces in the $2^n=2^9$ dimensional ambient space, which is achievable
as $10\cdot 2^5< 2^9$. 

More generally, in a sense degeneracy allows distinct errors to share the same 
orthogonal subspace and thus pack more efficiently. 
It must be pointed out though that this better packing is attained at the cost of $r$
gauge qudits compared to a stabilizer code.

In fact there exists another code among the Bacon-Shor codes which also 
beats the Hamming bound for the subsystem codes. This is the $[[16,1,9,4]]_2$ code.
%
The family of codes given in corollary~\ref{co:rectLattice} provides us 
with $[[12,1,6,3]]_2$, yet another example of a code
that beats the quantum Hamming bound like the $[[9,1,4,3]]_2$ code. 
 We can check that 
 $$ 
 \sum_{j=0}^1 \binom{12}{j}3^j = 37 > 2^{12-1-6}=2^5=32.
 $$
But note that unlike $[[9,1,4,3]]_2$ this code does not meet the Singleton bound for 
pure subsystem codes  as $6+1< 12-6+2$.
Naturally we can ask if there is a systematic method to construct codes that
beat the quantum Hamming bound. At the moment we do not know. It appears unlikely that
there exist long codes that beat the quantum Hamming bound. 

\section{Conclusion}\label{sec:conc}
We have proved that any $\F_q$-linear $[[n,k,r,d]]_q$ Clifford
subsystem code obeys the Singleton bound $k+r\le n-2d+2$.
Furthermore, we have shown earlier that pure Clifford subsystem codes
satisfy this bound as well. Our results provide much evidence for the
conjecture that the Singleton bound holds for arbitrary subsystem
codes.

Pure Clifford subsystem codes obey the Hamming (or sphere packing)
bound. In this paper, we have shown the amazing fact that 
there exist impure Clifford subsystem codes beating
the Hamming bound. This is the first
illustration of a case when impure codes pack more efficiently than
their pure counterparts. One example of a code beating the Hamming
bound is provided by the $[[9,1,4,3]]_2$ Bacon-Shor code; this
remarkable example also illustrates the following noteworthy facts:
\begin{compactenum}[a)]
\item The $[[9,1,4,3]]_2$ code requires $9-1-4=4$ syndrome measurements just like
the perfect $[[5,1,3]]_2$ code. 
\item Since $k+r\leq n-2d+2$ for all prime alphabet codes, $[[9,1,4,3]]_2$ code is also 
an optimal  subsystem code. 
This is interesting because the underlying classical
codes are not MDS. In MDS stabilizer codes, the underlying classical codes are required
to be MDS codes. 
\item The Bacon-Shor code is also impure. So unlike MDS stabilizer codes which must be pure,
MDS subsystem codes can be impure.
\item The maximal length of a $q$-ary stabilizer MDS code is $2q^2-2$, \shortcite{ketkar06}
whereas for subsystem codes it is larger as the $[[9,1,4,3]]_2$ code
indicates.
\end{compactenum}
The implication of b)--d) is that optimal subsystem codes can be
derived from suboptimal classical codes, unlike stabilizer
codes. 

We conclude with a few open questions that seem worth investigating.
\begin{compactenum}[i)]
\item Do arbitrary $[[n,k,r,d]]_q$ subsystem codes also satisfy
$k+r\leq n-2d+2$?
\item Is the Hamming bound for subsystem codes obeyed asymptotically?
\item What is the maximal length of MDS subsystem codes?
\end{compactenum}
The second question is motivated by the fact that binary stabilizer codes obey the
quantum Hamming bound asymptotically, see~\shortciteA{ashikhmin99}.

\smallskip 
\textit{Acknowledgments.} Part of this paper was presented at the
 BIRS workshop Operator
Structures in Quantum Information Theory, Banff, Canada, 2007 and at the 
I$^2$Lab Workshop: Frontiers in Quantum and Biological Information
Processing, Orlando, Florida 2006. We
thank the organizers, particularly David Kribs, Mary Beth Ruskai and
Pawel Wocjan, for inviting us to these fruitful workshops.  This
research was supported by NSF CAREER award CCF 0347310 and NSF grant
CCF 0622201. 

\appendix{Syndrome measurement for nonbinary $\F_q$-linear codes}\label{sec:appendix} 

Decoding of nonbinary quantum codes has not been studied as well as
binary codes.  Encoding of $\F_q$-linear nonbinary quantum codes was
investigated in \shortciteA{grassl03}.  The authors
suggest that the decoder is simply the encoder running backwards.
While that maybe reasonable in quantum communication, it is not
preferable in the case of quantum computation.

Here we give a method that allows us to measure the syndrome for
$\F_q$-linear nonbinary quantum codes.  We also show that an
$\F_q$-linear $[[n,k,r,d]]_q$ code requires $n-k-r$ syndrome
measurements.  But first we need the definition of the following
nonbinary gates, see \citeA{grassl03}.
\begin{compactenum}[i)]
\item $X(a)\ket{x}=\ket{x+a}$ 
\item $Z(b)\ket{x}=\omega^{\tr_{q/p}(bx)} \ket{x}$, $\omega=e^{j2\pi/p}$
\item $M(c)\ket{x}=\ket{cx}, c\in \F_q^\times$
\item $F\ket{x}= \frac{1}{\sqrt{q}}\sum_{y\in \F_q} \omega^{\tr_{q/p}(xy)}\ket{y}$
\item $A\ket{x} \ket{y} = \ket{x}\ket{x+y}$
\end{compactenum}
Graphically, these gates are represented below.
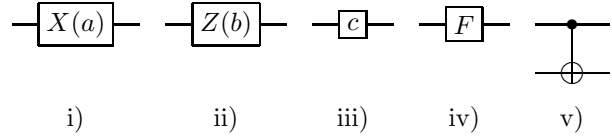
\[
\Qcircuit @C=1em @R=.7em {
& \gate{X(a)}&\qw && \gate{Z(b)}&\qw & &\gate{c} &\qw& &\gate{F}&\qw & &\ctrl{1} &\qw \\
& &  & & & & & &&&&&&\targ &\qw \\
& &  & & & & & &&&&&& &\\
& \text{i)}&  & &\text{ii)} & & &\text{iii)} &&&\text{iv)}&&&\text{v)} &
}
\]
Consider the following circuit. 
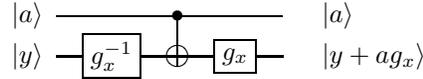
\[
\Qcircuit @C=1em @R=.7em {
\lstick{\ket{a}}& \qw & \ctrl{1} & \qw  &\qw &\rstick{\ket{a}}\\
\lstick{\ket{y}}& \gate{g_x^{-1}} & \targ{Z} & \gate{g_x} & \qw&\rstick{\ket{y+a g_x}}
}
\]
Alternatively, this circuit maps $\ket{a}\ket{x}$ to $\ket{a}X({a
g_x})\ket{y}$. Observe that this circuit effectively applies $X({a
g_x})$ on the second qudit.  Using the linearity, we can analyze the
following circuit.
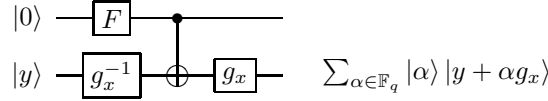
\[
\Qcircuit @C=1em @R=.7em {
\lstick{\ket{0}}& \gate{F}& \ctrl{1} & \qw  &\qw & \\ 
\lstick{\ket{y}}&\gate{g_x^{-1}} & \targ{Z} & \gate{g_x} & \qw&\rstick{\sum_{\alpha\in \F_q}\ket{\alpha}\ket{y+\alpha g_x}}
}
\]
The above circuit maps $\ket{0}\ket{y}$ to $\sum_{\alpha \in \F_q} \ket{\alpha} X({\alpha g_x})\ket{y}$. 
Using the fact that $F X(b) F^\dagger =Z(b)$, we can show that the following circuit maps 
$\ket{b}\ket{y}$ to $\ket{b} Z({bg_z}) \ket{y}$. 
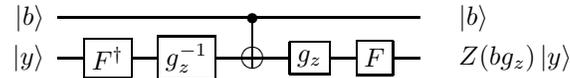
\[
\Qcircuit @C=1em @R=.7em {
\lstick{\ket{b}}& \qw&\qw & \ctrl{1} & \qw  &\qw &\qw&\rstick{\ket{b}}\\
\lstick{\ket{y}}& \gate{F^\dagger}&\gate{g_z^{-1}} & \targ{Z} & \gate{g_z} & \gate{F} &\qw&\rstick{Z({bg_z})\ket{y}}
}
\]
If we wanted to apply a general operator $X({ag_x})Z({ag_z})$ to the
second qudit conditioned on the first one, then we can combine the
previous circuits as follows.
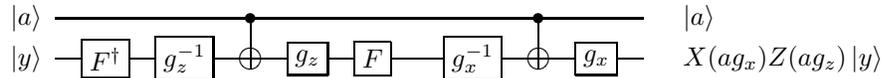
\[
\Qcircuit @C=1em @R=.7em {
\lstick{\ket{a}}& \qw&\qw & \ctrl{1} & \qw  &\qw  &\qw&\qw& \ctrl{1} & \qw  &\qw &\rstick{\ket{a}}\\
\lstick{\ket{y}}& \gate{F^\dagger}&\gate{g_z^{-1}} & \targ{Z} & \gate{g_z} & \gate{F} &\qw& \gate{g_x^{-1}} & \targ{Z} & \gate{g_x}&\qw &\rstick{X({ag_x})Z({ag_z})\ket{y}}
}
\]
The above implementation is not optimal in terms of gates, but it will
suffice for our purposes.  Consider an $[[n,k,r,d]]_q$ code. Let $E$
be an error in $\mathcal{E}$.  If $E$ is detectable, then $E$ does not
commute with some element(s) in the stabilizer of the code. Let
$$g=(g_x|g_z)=(0,\ldots,0,a_j,\ldots,a_n|0,\ldots,0,b_j,\ldots,
b_n)\in \F_q^{2n},$$ where $(a_j,b_j)\neq (0,0)$, be a generator of
the stabilizer. Then for all detectable errors that do not commute
with a multiple of $g$, the following circuit gives a nonzero value on
measurement.
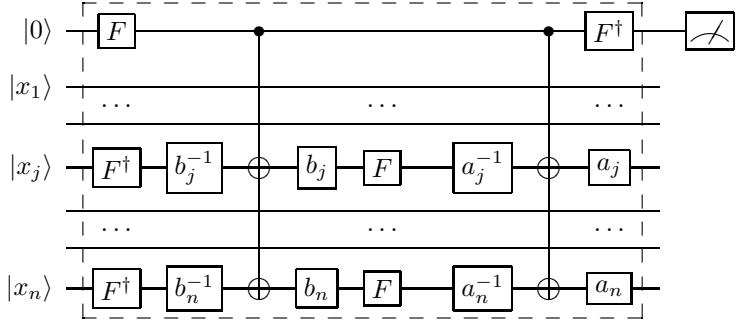
\[
\Qcircuit @C=1em @R=.7em {
\lstick{\ket{0}}& \gate{F}&\qw & \ctrl{9} & \qw  &\qw  &\qw&\qw& \ctrl{9} & \gate{F^\dagger}  &\qw &\meter\\
& & & &   &  &&& &  & &\\
\lstick{\ket{x_1}}& \qw&\qw & \qw & \qw  &\qw  &\qw&\qw& \qw & \qw  &\qw &\\
& \dots& & &   &\dots  &&& &  \dots &&\\
& \qw&\qw & \qw & \qw  &\qw  &\qw&\qw& \qw & \qw  &\qw &\\
\lstick{\ket{x_j}}& \gate{F^\dagger}&\gate{b_j^{-1}} & \targ{Z} & \gate{b_j} & \gate{F} &\qw& \gate{a_j^{-1}} & \targ{Z} & \gate{a_j}&\qw &\\
& \qw&\qw & \qw & \qw  &\qw  &\qw&\qw& \qw & \qw  &\qw &\\
& \dots& & &   &\dots  &&& & \dots & &\\
& \qw&\qw & \qw & \qw  &\qw  &\qw&\qw& \qw & \qw  &\qw &\\
\lstick{\ket{x_n}}& \gate{F^\dagger}&\gate{b_n^{-1}} & \targ{Z} & \gate{b_n} & \gate{F} &\qw& \gate{a_n^{-1}} & \targ{Z} & \gate{a_n}&\qw \gategroup{1}{2}{10}{10}{.7em}{--}
}
\]
Note that whenever $(a_i,b_i)=(0,0)$, then we leave that qudit alone. Similarly if $a_i$ or 
$b_i$ are zero, then we do not implement the corresponding portion. 
Let the input to the above circuit be $E\ket{\psi}$, where 
$\ket{\psi}$ is an encoded state. 
It can be easily verified that the above circuit maps the state $\ket{0}E\ket{\psi}$ to
$$ 
\sum_{\alpha \in \F_q}F^\dagger\ket{\alpha}X({\alpha g_x})Z({\alpha g_z})E\ket{\psi}.
$$
Let $X({g_x})Z({g_z})E=\omega^{\tr_{q/p}(t)}E X({g_x})Z({g_z})$, where $X({g_x})Z({g_z})$ 
is corresponding matrix representation of $g$. Then we have 
$X({\alpha g_x})Z({\alpha g_z})E=\omega^{\tr_{q/p}(\alpha t)}E X({g_x})Z({g_z})$, by 
lemma~5 in 
\cite{ketkar06}.
Thus we can write 
\begin{eqnarray*}
\sum_{\alpha\in \F_q}\ket{\alpha}  X({\alpha g_x})Z({\alpha g_z})E\ket{\psi} &= &\sum_{\alpha \in \F_q}\ket{\alpha}\omega^{\tr_{q/p}(\alpha t)} E  X({\alpha g_x})Z({\alpha g_z})  \ket{\psi},\\
&=&\left(\sum_{\alpha\in \F_q}\ket{\alpha} \omega^{\tr_{q/p}(\alpha t)} \right)E\ket{\psi},
\end{eqnarray*}
where we have made use of the fact that $X({\alpha g_x})Z({\alpha g_z}) \ket{\psi}=\ket{\psi}$
as  $X({\alpha g_x})Z({\alpha g_z})$ is in the stabilizer.
The final state is  given by 
\begin{eqnarray*}
\sum_{\alpha\in \F_q}F^\dagger\ket{\alpha}  X({\alpha g_x})Z({\alpha g_z})E\ket{\psi} &= &
\sum_{\alpha\in \F_q}F^\dagger\ket{\alpha} \omega^{\tr_{q/p}(\alpha t)} E\ket{\psi},\\
&=&\sum_{\alpha\in \F_q}\sum_{\beta\in \F_q}\omega^{-\tr_{q/p}(\alpha\beta)}\ket{\beta} \omega^{\tr_{q/p}(\alpha t)} E\ket{\psi},\\
&=&\sum_{\beta\in \F_q}\ket{\beta}\sum_{\alpha\in \F_q} \omega^{\tr_{q/p}(\alpha t -\alpha\beta )} E\ket{\psi},\\
&=&\sum_{\beta\in \F_q}\ket{\beta}\sum_{\alpha\in \F_q} \omega^{\tr_{q/p}(\alpha t -\alpha\beta )} E\ket{\psi},\\
&=&\ket{t}E\ket{\psi},
\end{eqnarray*}
where the last equality follows from the property of the characters of
$\F_q$.  Next we observe that the error $\alpha E$, where $\alpha\in
\F_q$ gives $\ket{\alpha t}$ on measurement. Strictly speaking we refer to
the preimage of $\alpha\ol{E}$ in $\mathcal{E}$.  Hence the syndrome
qudit can take $q$ different values. Since every detectable error
does not commute with some $\F_q$-multiple of a stabilizer generator,
we have the following lemma on the necessary and sufficient number of syndrome 
measurements.
\begin{lemma}\label{lm:suffMeas}
Given an $\F_q$-linear $[[n,k,r,d]]_q$ Clifford subsystem code, 
$n-k-r$ syndrome measurements are required for decoding it completely.
\end{lemma}
\begin{proof}
Let $g$ be a generator of the stabilizer of the subsystem code. 
By theorem~\ref{th:oqecfq} and lemma~\ref{lm:xzBasis}, for every generator $g$ there 
exists at least one detectable error that does not commute with $g$ but commutes 
with all the other generators. This error can be detected only by 
measuring $g$. Thus we need to measure all the
generators of the stabilizer, equivalently $n-k-r$ syndrome
measurements must be performed.

Every correctable error takes the code space into a $q^{k+r}$-dimensional orthogonal subspace in 
the $q^n$-dimensional ambient space, see \S\ref{sec:background}. Each of these errors will give a distinct syndrome. 
This implies that we can have $q^{n-k-r}$ distinct
syndromes. Since each  syndrome measurement can have $q$ possible outcomes and there are 
$n-k-r$ generators, these measurements are sufficient for performing error correction.
\end{proof}

This parallels the classical case where
an $[n,k,d]_q$ code requires $n-k$ syndrome bits. 
A subtle caveat must be issued to the reader. If we choose to perform 
bounded distance decoding, then it maybe possible that the set of
correctable errors can be distinguished by a smaller number of syndrome
measurements. But even in the case of (classical) bounded distance decoding it is
often the case that we need to measure all the syndrome bits. 


\begin{thebibliography}{}

\bibitem[\protect\citeauthoryear{%
Aly%
, Klappenecker%
\BCBL{}\ \BBA{} Sarvepalli%
}{%
Aly%
\ \protect\BOthers{.}}{%
{\protect\APACyear{2006}}%
}]{%
aly06}%
\APACinsertmetastar{%
aly06}%
Aly, S.~A.%
, Klappenecker, A.%
\BCBL{}\ \BBA{} Sarvepalli, P.~K.%
%
\newblock{}\BBOP{}2006\BBCP{}.
\newblock{}\BBOQ{}Subsystem codes.\BBCQ{}
\newblock{}\BIn{} \Bem{Forty-fourth annual {A}llerton conference on
  communication, control, and computing, {I}llinois, {USA}.}
\newblock{}
\newblock{}(eprint:quant-ph/0610153)

\bibitem[\protect\citeauthoryear{%
Ashikhmin%
\ \BBA{} Litsyn%
}{%
Ashikhmin%
\ \BBA{} Litsyn%
}{%
{\protect\APACyear{1999}}%
}]{%
ashikhmin99}%
\APACinsertmetastar{%
ashikhmin99}%
Ashikhmin, A.%
\BCBT{}\ \BBA{} Litsyn, S.%
%
\newblock{}\BBOP{}1999\BBCP{}.
\newblock{}\BBOQ{}Upper bounds on the size of quantum codes.\BBCQ{}
\newblock{}\Bem{IEEE Trans. Inform. Theory}, \Bem{4}, 1206-1216.

\bibitem[\protect\citeauthoryear{%
Bacon%
}{%
Bacon%
}{%
{\protect\APACyear{2006}}%
}]{%
bacon06a}%
\APACinsertmetastar{%
bacon06a}%
Bacon, D.%
%
\newblock{}\BBOP{}2006\BBCP{}.
\newblock{}\BBOQ{}Operator quantum error correcting subsystems for
  self-correcting quantum memories.\BBCQ{}
\newblock{}\Bem{Phys. Rev.~A}, \Bem{\bf 73}(012340).

\bibitem[\protect\citeauthoryear{%
Bacon%
\ \BBA{} Casaccino%
}{%
Bacon%
\ \BBA{} Casaccino%
}{%
{\protect\APACyear{2006}}%
}]{%
bacon06b}%
\APACinsertmetastar{%
bacon06b}%
Bacon, D.%
\BCBT{}\ \BBA{} Casaccino, A.%
%
\newblock{}\BBOP{}2006\BBCP{}.
\newblock{}\BBOQ{}Quantum error correcting subsystem codes from two classical
  linear codes.\BBCQ{}
\newblock{}\BIn{} \Bem{Forty-fourth annual {A}llerton conference on
  communication, control, and computing, {I}llinois, {USA}.}
\newblock{}

\bibitem[\protect\citeauthoryear{%
Grassl%
\ \BBA{} Beth%
}{%
Grassl%
\ \BBA{} Beth%
}{%
{\protect\APACyear{1996}}%
}]{%
grassl96}%
\APACinsertmetastar{%
grassl96}%
Grassl, M.%
\BCBT{}\ \BBA{} Beth, T.%
%
\newblock{}\BBOP{}1996\BBCP{}.
\newblock{}\BBOQ{}Improved decoding of quantum error correcting codes from
  classical codes.\BBCQ{}
\newblock{}\BIn{} \Bem{Proceedings of the workshop on {P}hysics and
  {C}omputation, {B}oston, {USA}}\ (\BPGS\ 28--31).
\newblock{}

\bibitem[\protect\citeauthoryear{%
Grassl%
, R{\"o}tteler%
\BCBL{}\ \BBA{} Beth%
}{%
Grassl%
\ \protect\BOthers{.}}{%
{\protect\APACyear{2003}}%
}]{%
grassl03}%
\APACinsertmetastar{%
grassl03}%
Grassl, M.%
, R{\"o}tteler, M.%
\BCBL{}\ \BBA{} Beth, T.%
%
\newblock{}\BBOP{}2003\BBCP{}.
\newblock{}\BBOQ{}Efficient quantum circuits for non-qubit quantum
  error-correcting codes.\BBCQ{}
\newblock{}\Bem{Internat. J. Found. Comput. Sci.}, \Bem{\bf 14}(5), 757--775.

\bibitem[\protect\citeauthoryear{%
Ketkar%
, Klappenecker%
, Kumar%
\BCBL{}\ \BBA{} Sarvepalli%
}{%
Ketkar%
\ \protect\BOthers{.}}{%
{\protect\APACyear{2006}}%
}]{%
ketkar06}%
\APACinsertmetastar{%
ketkar06}%
Ketkar, A.%
, Klappenecker, A.%
, Kumar, S.%
\BCBL{}\ \BBA{} Sarvepalli, P.~K.%
%
\newblock{}\BBOP{}2006\BBCP{}.
\newblock{}\BBOQ{}Nonbinary stabilizer codes over finite fields.\BBCQ{}
\newblock{}\Bem{IEEE Trans. Inform. Theory}, \Bem{\bf 52}(11), 4892--4914.
\newblock{}(eprint:quant-ph/0508070)

\bibitem[\protect\citeauthoryear{%
Klappenecker%
\ \BBA{} Sarvepalli%
}{%
Klappenecker%
\ \BBA{} Sarvepalli%
}{%
{\protect\APACyear{2006}}%
}]{%
pre7}%
\APACinsertmetastar{%
pre7}%
Klappenecker, A.%
\BCBT{}\ \BBA{} Sarvepalli, P.~K.%
%
\newblock{}\BBOP{}2006\BBCP{}.
\newblock{}\Bem{Clifford code constructions of operator quantum
  error-correcting codes.}
\newblock{}(eprint:quant-ph/0604161)

\bibitem[\protect\citeauthoryear{%
Knill%
}{%
Knill%
}{%
{\protect\APACyear{2006}}%
}]{%
knill06}%
\APACinsertmetastar{%
knill06}%
Knill, E.%
%
\newblock{}\BBOP{}2006\BBCP{}.
\newblock{}\Bem{On protected realizations of quantum information.}
\newblock{}(eprint: quant-ph/0603252)

\bibitem[\protect\citeauthoryear{%
Kribs%
}{%
Kribs%
}{%
{\protect\APACyear{2006}}%
}]{%
kribs06b}%
\APACinsertmetastar{%
kribs06b}%
Kribs, D.~W.%
%
\newblock{}\BBOP{}2006\BBCP{}.
\newblock{}\BBOQ{}A brief introduction to operator quantum error
  correction.\BBCQ{}
\newblock{}\Bem{Contemporary Mathematics}, \Bem{\bf 414}, 27--34.

\bibitem[\protect\citeauthoryear{%
Kribs%
, Laflamme%
\BCBL{}\ \BBA{} Poulin%
}{%
Kribs%
\ \protect\BOthers{.}}{%
{\protect\APACyear{2005}}%
}]{%
kribs05}%
\APACinsertmetastar{%
kribs05}%
Kribs, D.~W.%
, Laflamme, R.%
\BCBL{}\ \BBA{} Poulin, D.%
%
\newblock{}\BBOP{}2005\BBCP{}.
\newblock{}\BBOQ{}A unified and generalized approach to quantum error
  correction.\BBCQ{}
\newblock{}\Bem{Phys. Rev. Lett.}, \Bem{\bf 94}(180501).

\bibitem[\protect\citeauthoryear{%
Kribs%
, Laflamme%
, Poulin%
\BCBL{}\ \BBA{} Lesosky%
}{%
Kribs%
\ \protect\BOthers{.}}{%
{\protect\APACyear{2006}}%
}]{%
kribs06}%
\APACinsertmetastar{%
kribs06}%
Kribs, D.~W.%
, Laflamme, R.%
, Poulin, D.%
\BCBL{}\ \BBA{} Lesosky, M.%
%
\newblock{}\BBOP{}2006\BBCP{}.
\newblock{}\BBOQ{}Operator quantum error correction.\BBCQ{}
\newblock{}\Bem{Quantum Information \& Computation}, \Bem{\bf 6}, 382--399.

\bibitem[\protect\citeauthoryear{%
Poulin%
}{%
Poulin%
}{%
{\protect\APACyear{2005}}%
}]{%
poulin05}%
\APACinsertmetastar{%
poulin05}%
Poulin, D.%
%
\newblock{}\BBOP{}2005\BBCP{}.
\newblock{}\BBOQ{}Stabilizer formalism for operator quantum error
  correction.\BBCQ{}
\newblock{}\Bem{Phys. Rev. Lett.}, \Bem{\bf 95}(230504).

\end{thebibliography}

\def\cprime{$'$}

\label{lastpage}
\end{document}